\documentclass{article}
\usepackage{arxiv}

\usepackage{graphicx}
\usepackage{cite}
\usepackage{amsmath}
\usepackage{url}
\usepackage{color}
\usepackage{multirow}
\usepackage{array}

\usepackage{algorithm}

\usepackage{algpseudocode}

\begin{document}
%

\title{HyperLogLog (HLL) Security: Inflating Cardinality Estimates}

%
%
%


\author{
  Pedro Reviriego \\
  Universidad Carlos III de Madrid \\ 
  Legan\'es 28911, Madrid, Spain \\ 
  email: {\tt\small revirieg@it.uc3m.es}  \\
  \And
  Pablo Adell \\
  Universidad Carlos III de Madrid \\ 
  Legan\'es 28911, Madrid, Spain \\ 
  email: {\tt\small 100346194@alumnos.uc3m.es}  \\
   \And
  Daniel Ting \\ 
  Tableau Software \\ 
  Seattle, Washington, USA \\
  email: {\tt\small dting@tableau.com} \\
}

\maketitle


\begin{abstract}

Counting the number of distinct elements on a set is needed in many applications, for example to track the number of unique users in Internet services or the number of distinct flows on a network. In many cases, an estimate rather than the exact value is sufficient and thus many algorithms for cardinality estimation that significantly reduce the memory and computation requirements have been proposed. Among them, Hyperloglog has been widely adopted in both software and hardware implementations. The security of Hyperloglog has been recently studied showing that an attacker can create a set of elements that produces a cardinality estimate that is much smaller than the real cardinality of the set. This set can be used for example to evade detection systems that use Hyperloglog. In this paper, the security of Hyperloglog is considered from the opposite angle: the attacker wants to create a small set that when inserted on the Hyperloglog produces a large cardinality estimate. This set can be used to trigger false alarms in detection systems that use Hyperloglog but more interestingly, it can be potentially used to inflate the visits to websites or the number of hits of online advertisements. Our analysis shows that an attacker can create a set with a number of elements equal to the number of registers used in the Hyperloglog implementation that produces any arbitrary cardinality estimate. This has been validated in two commercial implementations of Hyperloglog: Presto and Redis. Based on those results, we also consider the protection of Hyperloglog against such an attack.     

\end{abstract}


%

\section{Introduction}

Count distinct or cardinality estimation is widely used in many computing and networking applications \cite{Cardinality}. For example, detecting changes in the number of connections or active nodes on a network may be an indication of a cyber-attack or a network scanning \cite{Cardinality_Detection_Infocom},\cite{HLL_Switches}. Similarly, it is useful to identify nodes that have many active connections on a network \cite{Supernodes-TDSC} and cardinality estimation is also useful to detect the spreading of worms \cite{Worm_cardinality}. In computing, cardinality estimation is also widely used for example to estimate the number of distinct users on a website and it is supported in most data processing libraries and databases \cite{Apache},\cite{RedisBloom},\cite{Presto}. 

An exact count of the number of distinct elements on a large set is costly and needs memory that is linear on the cardinality of the set. As in many applications, a reasonably accurate estimate is enough, many cardinality estimation algorithms have been proposed among which Hyperloglog \cite{HLL} is widely used \cite{Apache},\cite{RedisBloom},\cite{Presto},\cite{HLL_Google}. Hyperloglog uses a small number of registers, each having only a few bits to accurately estimate any practical cardinality value. This means that cardinality estimation can be implemented with a small memory footprint. Additionally, the operations to insert elements on the Hyperloglog are simple and only one memory access is needed per insertion. 

As Hyperloglog is used today in many systems, its privacy and security become important and have been recently studied \cite{Cardinality_Privacy},\cite{HLL_Attack}. In particular, for the security, it has been shown that an attacker that has no knowledge of the Hyperloglog implementation details can create a set of elements with large cardinality for which Hyperloglog produces a much smaller estimate. This could be used by an attacker to evade Hyperloglog based detection when performing for example a denial of service attack or network scan \cite{Cardinality_Detection_Infocom},\cite{HLL_Switches}. 

In this paper, Hyperloglog's security is studied from the opposite angle: an attacker that wants to create a set of elements that when inserted on the Hyperloglog will produce an estimate that is much larger than the number of elements in the set. This set can then be used by the attacker to for example trigger false alarms on Hyperloglog based detection systems but more importantly, also to inflate the estimates of the number of users, number of hits and similar performance metrics used for Internet services when estimated using Hyperloglog. The analysis shows that the attacker can create a set with the same number of elements as registers in the Hyperloglog (typically at most a few thousands) that produces any arbitrarily large cardinality value on the Hyperloglog cardinality estimation. This has been verified on two commercial implementations of Hyperloglog \cite{RedisBloom},\cite{Presto}. This is done without any knowledge of the Hyperloglog's implementation, which is worrying given the wide adoption of Hyperloglog. Fortunately, the protection scheme proposed in \cite{HLL_Attack} to detect attacks that reduced the Hyperloglog estimate can also be used to protect against this new attack that inflates the Hyperloglog estimate. 

The rest of the paper is organized as follows: the next section briefly discusses the Hyperloglog algorithm. Then, in section \ref{Sec_Inflating} the adversarial model and the proposed attack are presented and the detection of the attack is also briefly discussed. The attack is evaluated in section \ref{Sec_Evaluation} and the paper ends with the conclusion in section \ref{Sec_Conclusion}. 

\section{HyperLogLog}

As discussed in the introduction, HyperLogLog (HLL) is a data sketch that efficiently estimates number of unique elements in a data stream \cite{HLL}. It is formed by an array of $R$ registers $r_1, \ldots, r_R$ each having a small number of bits, typically four or five. To insert an element $x$ in the sketch, it is first mapped to one register $r_{h(x)}$ using a hash function $h(x)$. A second hash function $g(x)$ is used to generate a value $v(x)$ that is computed as one plus the number of leading zeros in the bit representation of $g(x)$. If $v(x)$ is larger than the value stored in register $r_{h(x)}$, then it is written on the register such that each register keeps the maximum value for all elements mapped to that register.

\begin{figure}[h]
  \centering
  \includegraphics[scale=0.4]{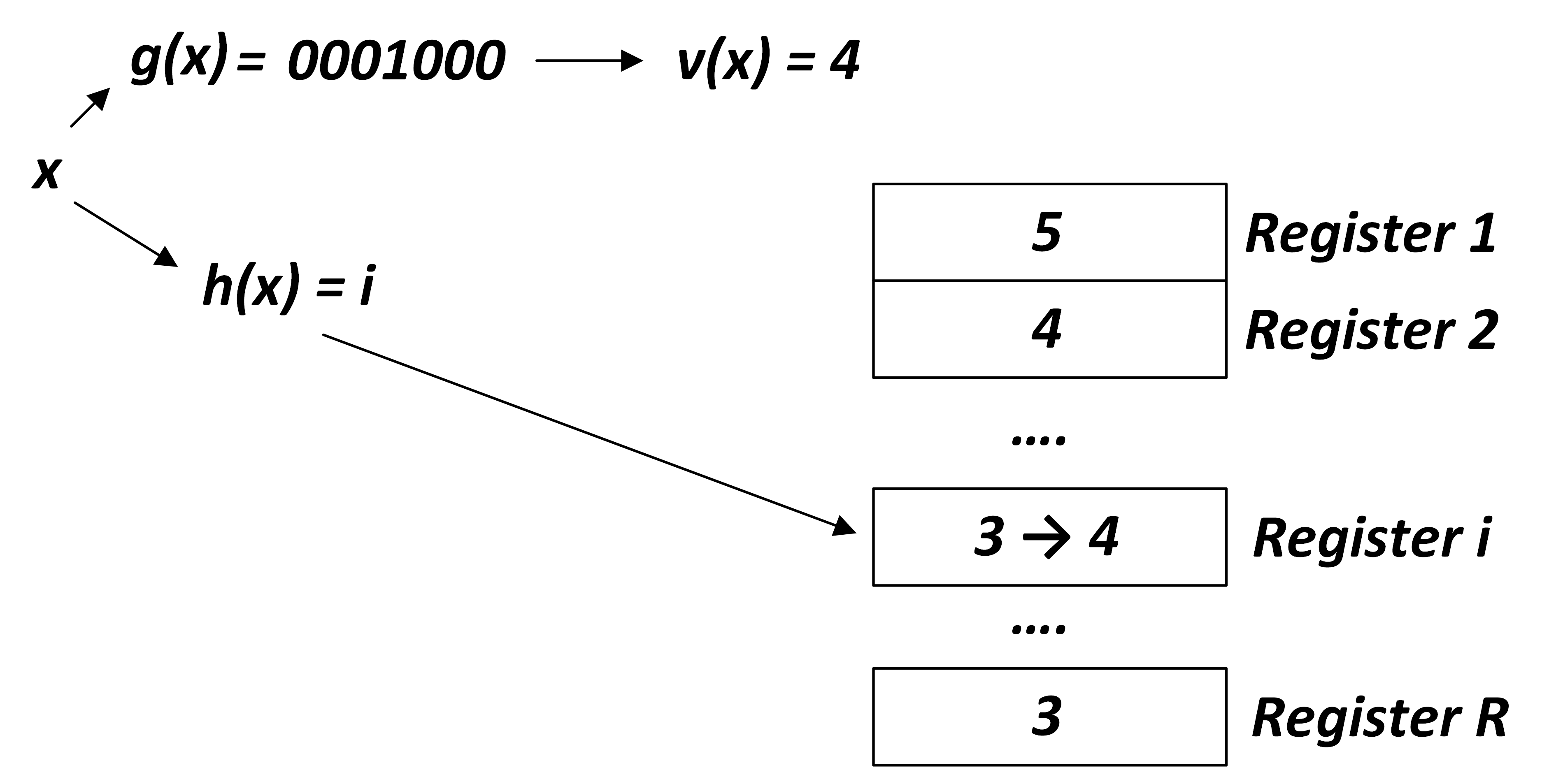}\\
  \caption{Example of a register update in HLL}\label{FigHLL2}
\end{figure}

Figure \ref{FigHLL2} illustrates an Hyperloglog showing the insertion of element $x$ that is mapped to register $i$ and has a value $v(x) = 4$. Since the register stores a value of 3 that is smaller, the new value is written to the register as shown in the Figure.  The probability of a register having a value with $t_z$ leading zeros is related to the number of elements mapped to the counter in an exponential way. This enables the registers to cover a wide range of cardinalities using only a few bits.     

To estimate the cardinality, Hyperloglog first computes the inverse of the harmonic mean of the registers contents as follows:

\begin{equation}
  Z = \sum_{i=1}^{R} 2^{-r_i}, 
\end{equation}

\noindent where $r_i$ is the value of the $i^{th}$ register.   

Then by multiplying by a constant factor given in \cite{HLL} that depends on the number of counters $R$, the cardinality estimate is obtained as: 

\begin{equation}
\label{eqn:harmonic mean}
  C_{HLL} = {\alpha}_R \frac{R^2}{Z} 
\end{equation}

For cardinalities that are much larger than the number of registers $R$, the relative error of the HLL estimate is approximately $\frac{1.04 \cdot C}{\sqrt{R}}$ where $C$ is the true cardinality. Therefore, by selecting the number of registers $R$, the desired accuracy can be obtained. For example with 1024 registers, the relative error is approximately 3\%. When the cardinality $C$ is comparable to $R$ or smaller, the estimate provided with equation \ref{eqn:harmonic mean} is worse and thus alternative estimates are used. The original Hyperloglog algorithm \cite{HLL} proposed using each register as a bit of a Linear Probabilistic Counting Array (LPCA) \cite{LPCA}. In more detail, the bit is one when the value of the register is different from zero and zero when it is zero.    Recently, more sophisticated estimates for low cardinalities have been proposed \cite{LowRange_ERTL}.

Hyperloglog sketches that use the same hash functions $g(x)$ and $h(x)$ can be merged to compute the cardinality of the union of their data streams. To do so, a new sketch is created by taking for each register the maximum of values of the same register across all the merged sketches. This merging is useful in many applications.

Before moving to the next section, an interesting observation is that once all elements in the data stream have been inserted, the set $X = {x_1,x_2,...,x_R}$ formed by $R$ elements each mapping to one register such that $x_i$ maps to register $i$ and has value $v(x_i)= r_i(x)$ would produce the same Hyperloglog estimation as the entire data stream.

\section{Inflating HyperLogLog Cardinality Estimates}
\label{Sec_Inflating}

This section first presents the model considered for the attacker. Then the proposed attacked is presented and finally the protection against such an attack is briefly discussed. 

\subsection{Adversarial Model}

The assumption is that the attacker can create Hyperloglog instances that use the same hash functions as the target Hyperloglog instance being attacked, perform insertions, and check the Hyperloglog estimate. Using the same hash functions is critical in many applications of Hyperloglog as it allows Hyperloglog sketches to be merged to compute cardinalities of unions. On the other hand, the attacker has no information on the specific implementation of Hyperloglog. In particular the hash functions $g$ and $h$ and the number of registers $R$ are not known. Information about an item $x$'s hash values $g(x), h(x)$ can only be obtained indirectly through observing cardinality estimates before and after inserting $x$ into a Hyperloglog instance. This attacker model is the same as the one considered in our previous work \cite{HLL_Attack}.

\subsection{Inflating Hyperloglog (HLL) Estimates}

Let us consider a set of elements $S$ with cardinality $C$. After inserting the elements in $S$ on an HLL with $R$ registers, by construction of the HLL instance, for each register $i$ there is an element $x_i$ that maps to the register and has its final value $v(x_i) = r_i(x)$. Therefore, if we can identify those $R$ elements $X = {x_1,x_2,...,x_R}$, by inserting them on the HLL, the cardinality estimate would be $C$ instead of $R$. This means that the cardinality estimate of HLL can be set to any arbitrary value $C$ using a set of only $R$ elements. However, finding those $R$ elements without having any details on how the HLL is implemented does not seem to be straightforward. 

To find those $R$ elements, we first insert the elements in set $S$ on an empty HLL and after inserting each element $y$ we check if it has increased the HLL estimate and if so we add it to an initial set $Y$. This set will in general be different from $X$. This is due on one hand to the fact that when inserting elements from $S$ on register $i$, the register can take several intermediate values until reaching its final value $r_i$. Then, the elements that correspond to those intermediate values will be inserted on set $Y$ if they increment the HLL estimate. On the other hand, if the cardinality estimates are rounded to the nearest integer, there may be cases in which inserting the element $x_i$ that sets the final value $r_i$ does not increment the HLL estimate and thus $x_i$ would not be added to $Y$. This can occur if the HLL implementation uses the LPCA estimate for low cardinalities as discussed in the previous section. Then when few elements have been inserted if there is an element already mapped to register $i$, inserting $x_i$ will not increment the LPCA estimate (as the LPCA bit was already a one after the first element was inserted) and thus $x_i$ will not be added to $Y$. When more elements have been inserted into the HLL instance so that the HLL estimate is used, if the register has a value that is large relative to the number of elements that have been inserted so far, incrementing its value may also not increase the HLL estimate and thus again $x_i$ may also not be inserted on $Y$. Therefore, this initial set $Y$ will have both elements that are not in $X$ and miss elements from $X$. An initial analysis of the impact of the missing elements on the HLL cardinality estimate is presented in the Appendix. A more detailed analysis is left for future work. 

To illustrate those cases, let us consider an HLL with $R=1024$ registers an a set $S$ with a cardinality $C = 100,000$. Then on average approximately 100 elements would map to each register. Let us focus on register $i$ and assume that there are two elements $a$ and $b$ that map to that register with values $v(a) = 3$ and  $v(b) = 4$ and that both are the first two elements inserted on the HLL. Then, if the HLL uses the LPCA estimate for low cardinalities, the estimate would depend only on the number of registers that are not zero and this will only change when inserting $a$ so that $b$ will not be added to $Y$. Now, let us consider another register $j$ to which also two elements $c$ and $d$ map with $v(c) = 7$ and  $v(d) = 8$ and that both are added when the cardinality estimate for HLL is already 5,000. Then the sum of the values of the HLL registers used in the denominator of equation \ref{eqn:harmonic mean} would be approximately 150 on average. Changing the value of register $j$ from 0 to 7 when inserting $c$ would decrease the average to approximately 149, thus increasing the cardinality by approximately $5000 / 149 \approx 34 > 1$. A subsequent insertion of $d$ would introduce a small change in the HLL estimate that is approximately $2^{-8}$ times smaller and thus lower than one and likely not detectable if estimates are rounded. Therefore, $d$ is not added to the set. Finally, let us consider a third register $k$ to which elements $e$ and $f$ map with $v(e) = 3$ and  $v(f) = 4$ and both are added when the HLL estimate is already close to 100,000. Then, both will increment the HLL estimate and thus will be added to the set $Y$. These three cases show how elements in $X$ may not be added to $Y$ and also how additional elements that are not in $X$ can be added to $Y$. However, the initial set $Y$ should produce an HLL estimate that is reasonably close to $C$ and thus can be used as an initial step to build set $X$.

Once set $Y$ has been constructed, to identify the missing elements from $X$, we can do the following. Build a new HLL and insert only the elements in $Y$. Then insert all the elements in $S$ and add those that increment the HLL estimate to $Y$. Now, since the cardinality estimate for $Y$ would be larger than the range for which the LPCA is used and close to $C$, the missing elements from $X$ should increase the cardinality estimate and thus would be added to $Y$.

Finally, to obtain the final set, we can build another empty HLL and insert the elements from $Y$ in reverse order. That is the elements that were added last to $Y$ are inserted first. After inserting each element, we check if the HLL estimate has increased and if so we add the element to a new set $V$. This procedure by construction will select for each register, the element in $Y$ that has the maximum value and thus set $V$ will tend to be the same or very similar to set $X$. The entire algorithm is summarized in Algorithm \ref{Alg01}. 

\begin{algorithm}
\small 
\caption{Procedure to generate the attack set $V$ } \label{Alg01}
{
\begin{algorithmic}[1]
 \State construct a set $S$ with the target cardinality $C$. 
 \State construct an empty HLL instance. 
 \State construct empty sets $Y$ and $V$. 
 \State \textit{---- PHASE 1: build the initial version of $Y$ ----} 
 \For{$s$ in $S$}
   \State get HLL estimate $HLL_{before}$.
   \State insert $s$ on the HLL.
   \State get HLL estimate $HLL_{after}$.
   \If {$HLL_{after} > HLL_{before}$}
      \State add $s$ to set $Y$.
   \EndIf 
 \EndFor
 \State \textit{---- PHASE 2: add missing elements to $Y$ ----} 
 \State construct an empty HLL instance. 
 \For{$y$ in $Y$}
   \State insert $y$ on the HLL.
 \EndFor
 \For{$s$ in $S$}
   \State get HLL estimate $HLL_{before}$.
   \State insert $s$ on the HLL.
   \State get HLL estimate $HLL_{after}$.
   \If {$HLL_{after} > HLL_{before}$}
      \State add $s$ to set $Y$.
   \EndIf 
 \EndFor
 \State \textit{---- PHASE 3: build $V$ from $Y$ ----} 
 \State construct an empty HLL instance. 
 \For{$y$ in $Y$ starting from last to first inserted}
   \State get HLL estimate $HLL_{before}$.
   \State insert $y$ on the HLL.
   \State get HLL estimate $HLL_{after}$.
   \If {$HLL_{after} > HLL_{before}$}
      \State add $y$ to set $V$.
   \EndIf 
 \EndFor
\end{algorithmic}
}
\end{algorithm}

From the description of the algorithm, it can be seen that the number of operations required on the Hyperloglog to build an attack set of $R$ elements that produces a cardinality estimate of $C$ is $O(C)$. That means that the complexity of the attack is of the same order to that of inserting a set of $C$ elements on the Hyperloglog and thus it would be feasible to build attack sets for common cardinality values.

\subsection{Protecting Against the Attack}

The attack described in the previous subsection poses a challenge to Hyperloglog security and thus it is important to discuss what can be done to detect and prevent the attack. The use of a random salt in the hash functions $h(x)$ and $g(x)$  so that each Hyperloglog instance is different is one alternative, but as discussed in \cite{HLL_Attack}, it means that Hyperloglog sketches can no longer be merged, which is a key feature for many applications. An alternative to detect the attack could be the use of two Hyperloglog sketches: the main one not salted so that it can be merged and an auxiliary one salted as proposed in \cite{HLL_Attack}. Then, if the estimates of the two sketches are significantly different, an attack is detected. This Salted Not Salted (SNS) protection proposed in \cite{HLL_Attack} to detect attacks that try to reduce the Hyperloglog estimate can also detect the inflating attack described in the previous subsection. 

Simpler techniques may also be used to detect the attack. For example, logging the percentage of Hyperloglog insertions that increase the register values and the average increment of the value. In a normal state, only a small fraction of the insertions would increment the register values once the cardinality estimate is larger than $R$. The average increment on the value would also be small with one having a probability of approximately 50\%, two of 25\%, three of 12.5\% and so on giving an average of close to two. Therefore, if at some point either a large fraction of the insertions increment the value or the average increment is large, an attack is detected. The detailed study of these detection schemes is left for future work.

\section{Evaluation}
\label{Sec_Evaluation}

\begin{figure}[h]
  \centering
  \includegraphics[scale=0.6]{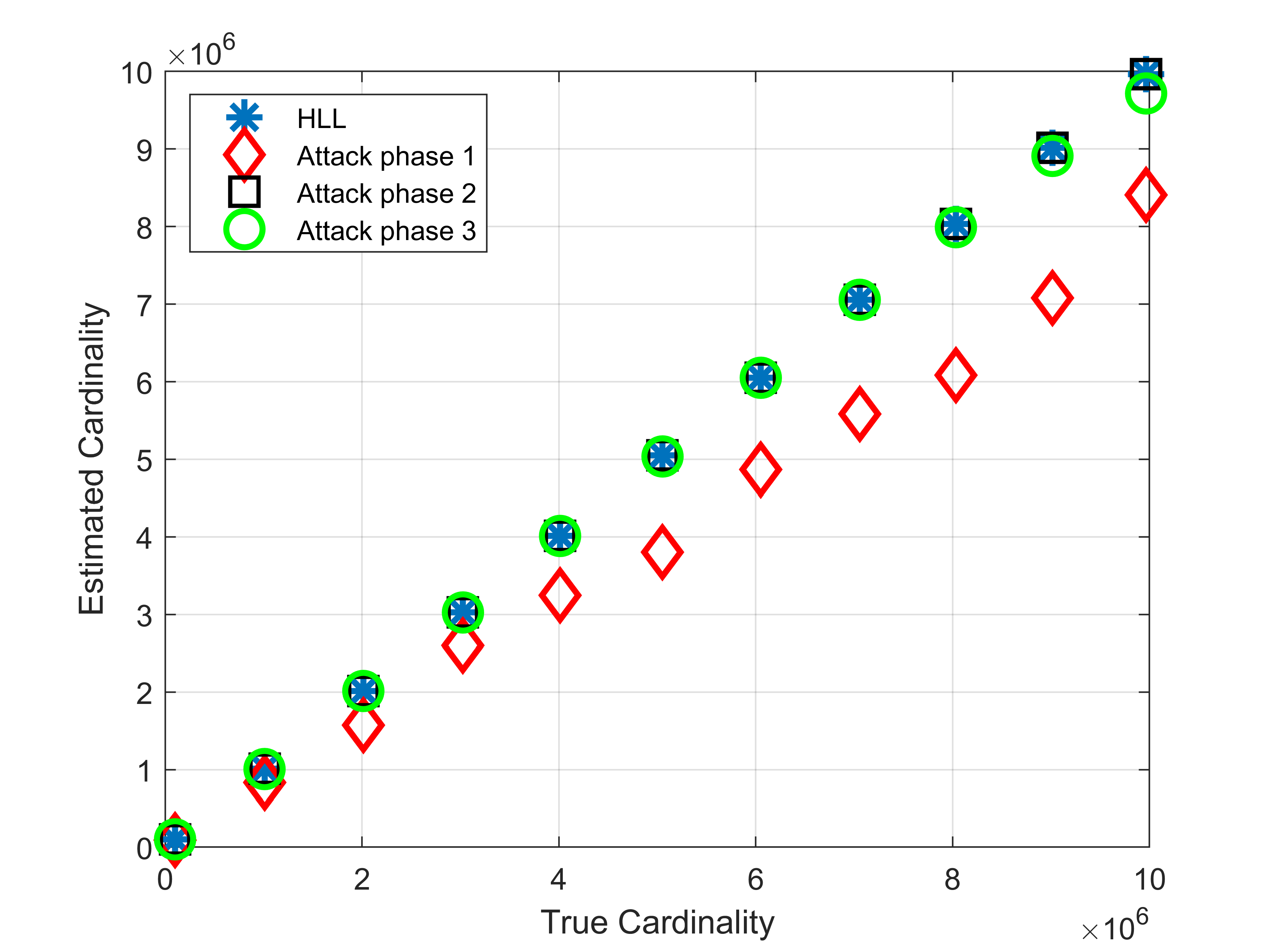}\\
  \caption{Hyperloglog Cardinality Estimate for the original set that the attack set in phases 1,2,3}\label{FigCard}
\end{figure}

\begin{figure}[h]
  \centering
  \includegraphics[scale=0.6]{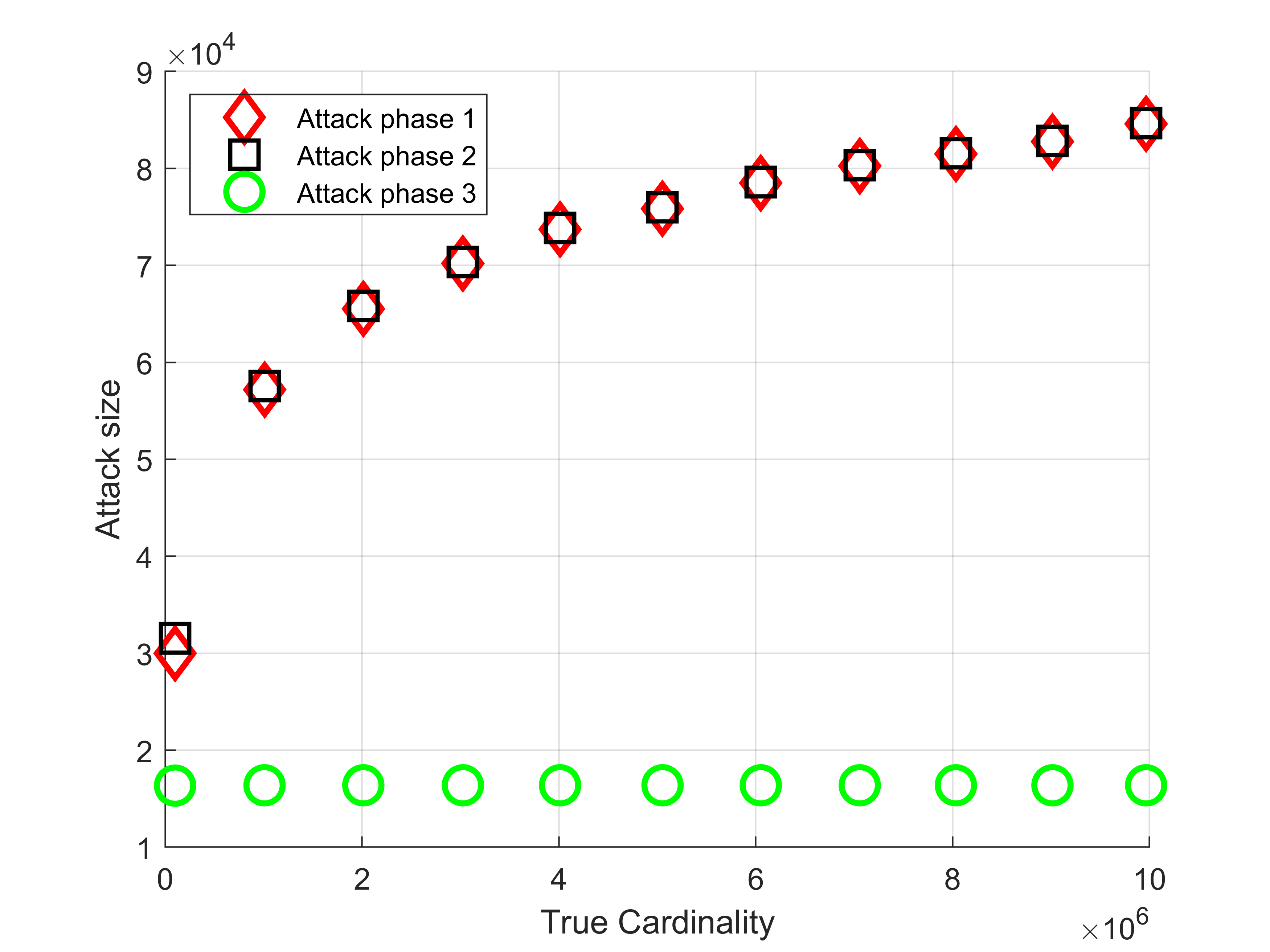}\\
  \caption{Size of the attack set in phases 1,2,3}\label{FigAttSize}
\end{figure}

The proposed attack has been implemented in Redis \cite{RedisBloom} that uses a default Hyperloglog implementation with $R=16384$ and in Presto \cite{Presto} that uses by default $R=4096$ registers\footnote{The code for the attack both for Redis and Presto is available in \url{https:// github.com/adell13pablo/hll_attack}}. For Redis, the estimated cardinalities for the attack set on phases 1,2,3 are shown in Figure \ref{FigCard} and the attack set sizes on Figure \ref{FigAttSize}. It can be seen that the attack achieves the same cardinality estimate as that of the entire set in phase 2 and phase 3. In phase 1, the attack set produces also a large cardinality estimate but smaller than that of the original set. This is inline with the expected results. Finally, the attack size is significantly reduced in phase 3 bringing very close to the value of $R$ again as expected. 

After validating the attack in Redis, it was also implemented in Presto \cite{Presto}. In this case, the default configuration of Hyperloglog uses $R = 4096$ registers. Then attack was run for sets with cardinalities of $C = 20,000, 40,000, 60,000, 80,000$ and $100,000$. The results are summarized in Tables \ref{TabPresto1},\ref{TabPresto2} that show the Hyperloglog cardinality estimate when inserting the attack set and the size of the attack set for the three phases of algorithm \ref{Alg01}. It can be seen how the algorithm increases the cardinality estimate in the second phase and reduces the size of the attack set in the third phase to obtain a final attack set of approximately $R= 4096$ elements that creates an Hyperloglog estimate of approximately $C$. These results are similar those of Redis and confirm the analysis presented in the previous section and the feasibility of the attack.


\begin{table}[h]
\caption{Cardinality estimate in Presto } 
\centering 
\begin{tabular}{ c c c c c c}  
\hline\hline 
Cardinality  & 20,000  & 40,000  & 60,000  & 80,000 & 100,000  \\  
\hline\hline 
Phase 1     & 17,128  & 36,756 & 48,332 & 63,940 & 78,372 \\ 
\hline
Phase 2     & 19,739 & 40,005  & 59,132  & 79,134 & 98,723 \\
\hline
Phase 3     & 19,739 & 39,625  & 59,132  & 79,314 & 98,723 \\
\hline

\end{tabular}
\label{TabPresto1} 
\end{table}

\begin{table}[h]
\caption{Attack set size in Presto } 
\centering 
\begin{tabular}{ c c c c c c}  
\hline\hline 
Cardinality  & 20,000  & 40,000  & 60,000  & 80,000 & 100,000  \\  
\hline\hline 
Phase 1     & 6883  & 9306 & 10,143 & 11,018 & 11,691 \\ 
\hline
Phase 2     & 7321 & 9715  & 10,396  & 11,234 & 11,870 \\
\hline
Phase 3     & 4088 & 4124  & 4125  & 4127  & 4134 \\
\hline

\end{tabular}
\label{TabPresto2} 
\end{table}

\section{Conclusion}
\label{Sec_Conclusion}

In this paper, the security of Hyperloglog against an attacker that wants to artificially inflate the cardinality estimate has been considered. The analysis shows that even without any knowledge of the Hyperloglog implementation, the attacker can create a small set of elements that when inserted on the Hyperloglog sketch produces a large cardinality estimate. The feasibility of the attack has been demonstrated by implementing and testing it on two commercial products that use Hyperloglog for cardinality estimation. The experimental results confirm the validity of the analysis and suggest that Hyperloglog implementations should incorporate mechanisms to detect and prevent such an attack.

The detection of the attack has also been briefly discussed showing that previous protection techniques are applicable to detect the attack. In addition, simpler techniques to detect the attack have also been identified and their detailed studied is left for future work. More generally, the security of data sketches should be carefully studied as computing systems rely increasingly on them to perform many operations beyond cardinality estimation.


\section*{Acknowledgment}

 Pedro Reviriego would like to acknowledge the support of the ACHILLES project PID2019-104207RB-I00 and the Go2Edge network RED2018-102585-T funded by the Spanish Ministry of Science and Innovation and of the Madrid Community research project TAPIR-CM  grant no. P2018/TCS-4496.
 
 \appendix
\section{Appendix}
 
 This appendix discusses in more detail first why some elements of the optimal attack set are not identified in the first phase of the attack algorithm and then why those missing elements have a non negligible impact on the cardinality estimate.    
 
 \subsection{Missing elements in Phase 1}
 
 There are two effects that make us miss elements when building the attack set on the first Phase. The first one is due to the use of the LPCA estimate when cardinality is low. For each register in the sketch, only the first element hashed to it affects the LPCA estimate. All subsequent elements in the register are ignored until the sketch switches to the HLL estimate.
 The switch between the LPCA and HLL estimates is typically done when the cardinality is $\approx 2.5\cdot R$ \cite{LowRange_ERTL}. 
 So there are in expectation 2.5 items per bin at the time of the switch. If there are $N$ distinct items in total, the number of missed items $M$ should be roughly 
 $M \approx R \cdot [2.5 / (N/R)] \cdot (1.5/2.5) = 1.5 R^2 / N$.
 That is $R$ times the chance that the largest hash value in a register is inserted during the LPCA phase of the sketch and is not the first one hashed to the register.
  
 The second effect that makes us miss elements is that some of them produce a small increment on the cardinality estimation. Therefore, if the HLL implementation returns an integer value, the increment may not be detected. For example, if the current estimate is 4002.1 and the increment is only 0.2 giving 4002.3, the HLL query would return 4002 in both cases and the element is missed. To see how this happens, let us start with the cardinality estimate is given by: 
 
\begin{equation}
    C_{est} = \alpha \cdot \frac{R^2}{Z}  
\end{equation}
 
a change on a counter from value $c_{old}$ to $c_{new}$ changes the value of $Z$ by:

\begin{equation}
\delta = 2^{-c_{old}}-2^{-c_{new}}
\end{equation}

and the estimated cardinality by:

\begin{equation}
I = \alpha \cdot \frac{R^2}{Z-\delta} -  \alpha \cdot \frac{R^2}{Z} =  \alpha \cdot \frac{R^2 \cdot \delta}{Z \cdot (Z-\delta)}
\end{equation}

that can be approximated by:

\begin{equation}
I \approx \alpha \cdot \frac{R^2 \cdot \delta}{Z^2} = \frac{\delta \cdot (C_{est})^2}{\alpha \cdot R^2}.
\end{equation}

The increment will be smaller than 0.5 when:
 
\begin{equation}
\delta < 0.5 \cdot \frac{\alpha \cdot R^2}{(C_{est})^2} 
\end{equation}

For example, if $R = 4096$ and $C_{est}= 20000$ values of $\delta$ lower than 0.015 may not be detected. For example, a counter that is incremented from 6 to 8 has a lower $\delta$ and may not be detected.  To analyze the probability that one of elements of the attack set is missed due to this second effect we can start by noting that $\delta < 2^{-c_{old}}$ and thus for an element to be missed:

\begin{equation}
2^{-c_{old}} < 0.5 \cdot \frac{\alpha \cdot R^2}{(C_{est})^2} 
\end{equation}

Taking the $log_2$ on both sides of the equation we get:

\begin{equation}
 c_{old} > 1+log_2(\alpha)+ 2 \cdot log_2(\frac{C_{est}}{R}) 
\end{equation}

This is much larger than the expected value of a register when the HLL has an estimate of $C_{est}$ that is $1+log_2(\frac{C_{est}}{R})$. Therefore, it seems unlikely that a register is in a state (has a large enough $c_{old}$ value) to miss an element. For example, if the HLL has $R=4096$ and $C_{est}= 32000$, the expected value would be around four and the condition to miss elements would require a value larger than approximately eight so few counters would meet that condition. As the cardinality grows, the probability would be even lower. Therefore, it seems that this second effect would have less impact than the first and is not considered further.

\subsection{Impact of the missing elements on the cardinality estimate}
 
The number of missed elements can be approximated by $1.5 R^2 / N$ as discussed in the previous subsection. It would seem that when $N$ is large (as it would be the case in the proposed attack as we are trying to inflate the HLL estimate), the value would be small. For example, if $R= 4096$ and $N = 1,000,000$, only 25 elements would be missed from the expected 4096 elements. However, although the number of missed elements is small the impact of these missed elements on the cardinality estimate is significant as was seen in the evaluation results. To understand why, an example can be used, consider a counter to which many elements map and there is one with maximum value $v_{max}$. If that element happens to be missed, then the rest of the elements will not increment the counter. Therefore, that counter for the attack set will take the value of the first element that mapped to it that will on average be low. This can significantly affect the cardinality estimate. For example, if $v_{max}= 10$ and the first element has a value of 1, the counter would take a much lower value.

The probability that this happens on a counter would be approximately $\frac{R}{C}$ as on average $\frac{C}{R}$ elements map to each counter thus the probability of picking the maximum is the inverse of that assuming $\frac{C}{R}$ is much larger than one. Therefore, on average this will occur for $\frac{R^2}{C}$ counters. For each of them, the value would be reduced to an average value of close to 2. Therefore the relative impact on the denominator of the HLL estimate could be roughly approximated by:

\begin{equation}
  Z_a = \frac{1}{\sum_{i=1}^{R} 2^{-r_i}+\frac{R^2}{C} \cdot (2^{-2}-\frac{R}{C})} 
\end{equation}

\begin{figure}[h]
  \centering
  \includegraphics[scale=0.6]{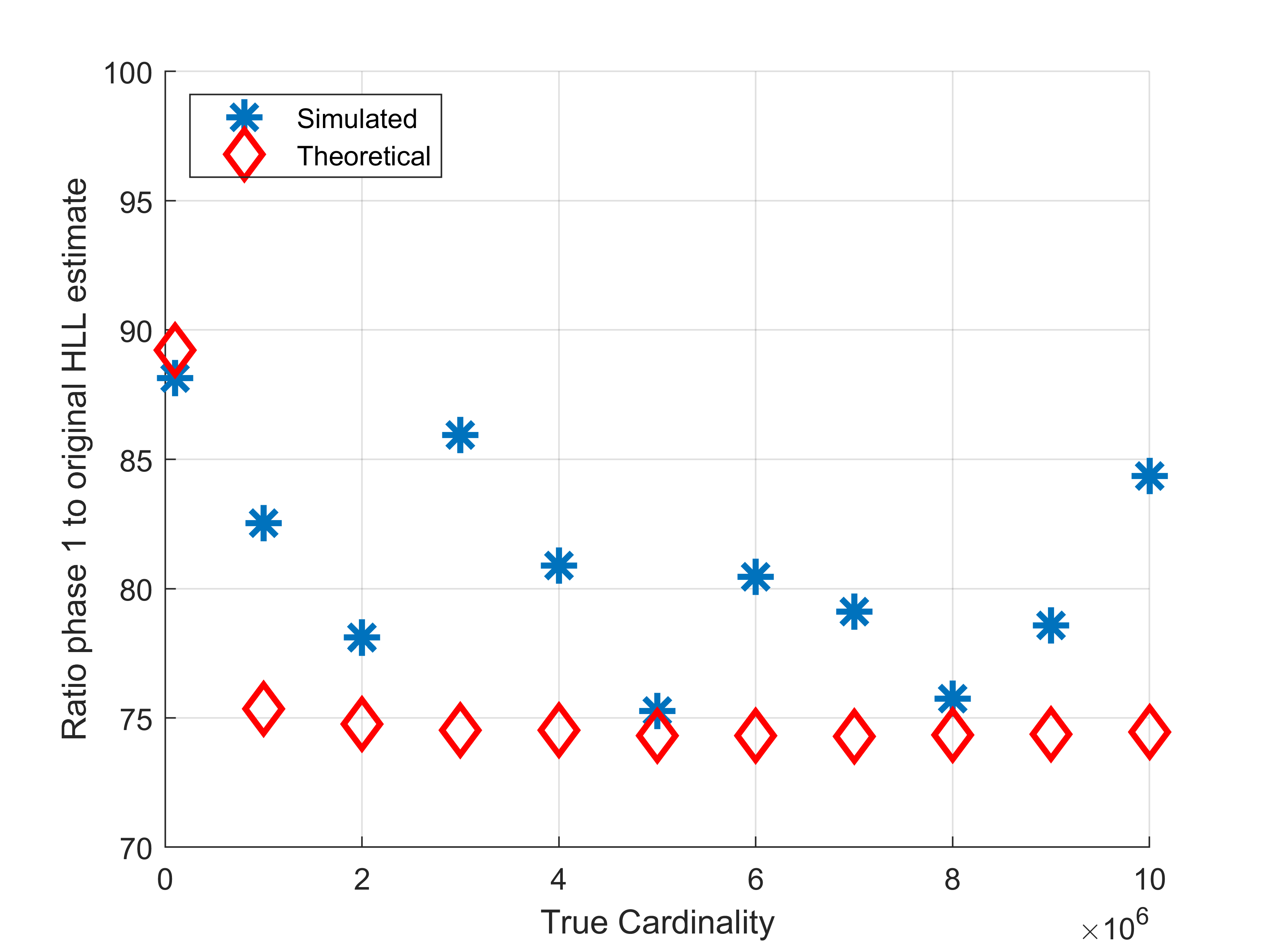}\\
  \caption{Ratio of the HLL estimate for the attack set in phase 1 to the HLL estimate for the original set}\label{FigEstCardRatio}
\end{figure}

The ratio of the cardinality of the attack set to that of the initial set is shown in Figure \ref{FigEstCardRatio} that also shows the rough theoretical estimate just presented. It can be seen that there is a lost of approximately 20 to 30\% and that the theoretical estimate is inline with the simulation results. This is because even a few counters can significantly affect the HLL estimate if they take values that are much lower than the rest of the counters as previously discussed in \cite{RM}.

\end{document}